\documentclass[prc,reprint,nofootinbib]{revtex4-1}
\usepackage{graphicx}   
\usepackage{latexsym}   

\begin{document}
\title{Heavy baryonic resonances, multistrange hadrons, and equilibration\\ at energies available at the GSI Schwerionensynchrotron, SIS18}

\author{J. Steinheimer$^1$, M. Lorenz$^3$, F. Becattini$^4$, R. Stock$^{1,3}$ and M. Bleicher$^{1,2}$}

\affiliation{$^1$ Frankfurt Institute for Advanced Studies, Ruth-Moufang-Str. 1, 60438 Frankfurt am Main, Germany}
\affiliation{$^2$ Institut f\"ur Theoretische Physik, Goethe Universit\"at Frankfurt, Max-von-Laue-Strasse 1, D-60438 Frankfurt am Main, Germany}
\affiliation{$^3$ Institut f\"ur Kernphysik, Goethe-Universit\"at, 60438 Frankfurt, Germany}
\affiliation{$^4$ Universita di Firenze and INFN Sezione di Firenze, Firenze, Italy}


\begin{abstract}
We study the details and time dependence of particle production in nuclear collisions at a fixed target beam energy of
$E_{\mathrm{lab}}= 1.76$ A GeV with the UrQMD transport model. We find that the previously proposed production mechanism for multi strange hadrons, $\phi$ and $\Xi$, are possible due to secondary interactions of incoming nuclei of the projectile and target with already created nuclear resonances, while the Fermi momenta of the nuclei play only a minor role. We also show how the centrality dependence of these particle multiplicities can be used to confirm the proposed mechanism, as it strongly depends on the number of participants in the reaction.
Furthermore we investigate the time dependence of particle production in collisions of Ca+Ca at this beam energy, in order
to understand the origins of the apparent chemical equilibration of the measured particle yields. We find that indeed 
the light hadron yields appear to be in equilibrium already from the very early stage of the collision while in fact no local equilibration, through multi collision processes, takes place. The apparent equilibrium is reached only through the decay of the resonances according to available phase space.
\end{abstract}

\maketitle

\section{Introduction}
The study of hadron production in nuclear collisions, at beam energies of a few GeV per nucleon, 
has the potential to give insights not only on the properties of hot and dense hadronic matter \cite{Randrup:1980qd,Koch:1986ud,Aichelin:1986ss},
but also on the fundamental mechanisms which govern particle production in these collisions.
In particular strange particles, which are produced below their elementary production threshold have proven
to show interesting properties. Here one tries to understand the large production
probability for multi strange hadrons, the $\phi$ and the $\Xi^-$, as measured by experiments at 
the SIS18 accelerator \cite{Agakishiev:2015xsa,Agakishiev:2009ar,Agakishiev:2009rr,Kolomeitsev:2012ij,Piasecki:2014zms}.
Furthermore the fact that most light hadron multiplicities in A+A and p+A collisions,
at these energies, can be well approximated by a canonical ensemble in chemical equilibrium \cite{Agakishiev:2015bwu,Agakishiev:2010rs} is yet to be understood microscopically.

\begin{figure*}[t]	
\includegraphics[width=\textwidth]{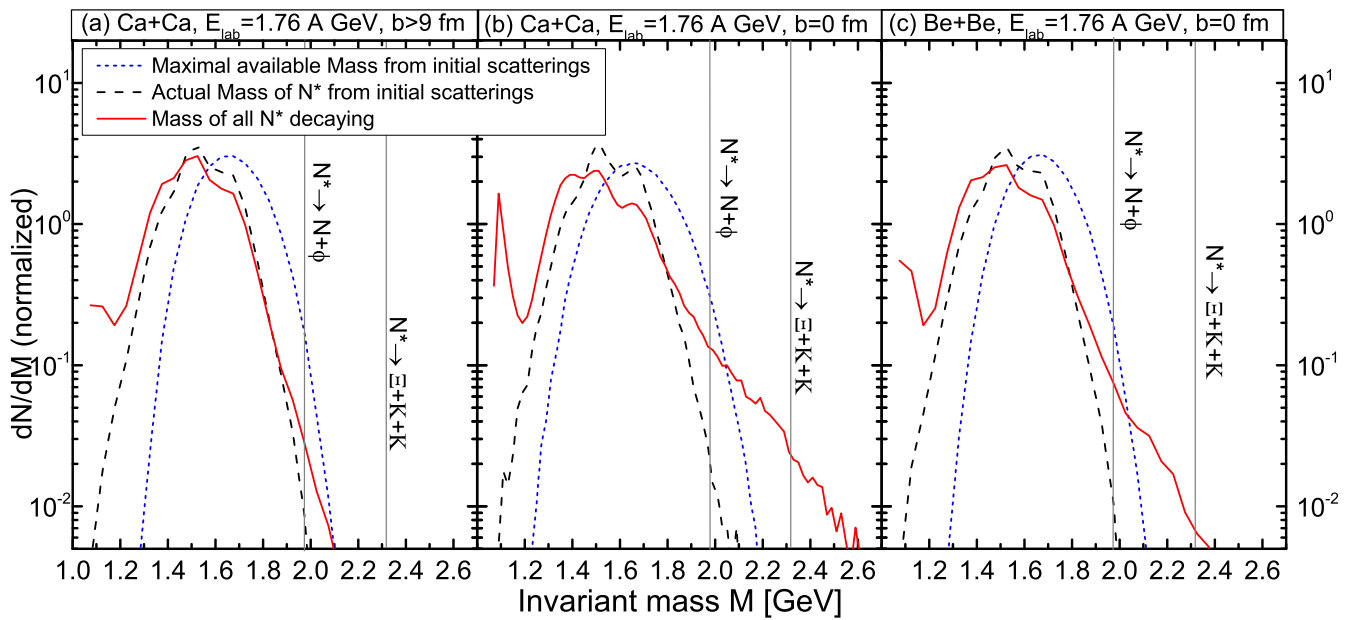}	
\caption{[Color online] Shown (as blue short dashed line) is the distribution of the maximally available invariant mass for resonance creation from initial scatterings of target and projectile nucleons. Note that the peak of the distribution corresponds to the energy available in proton+proton collisions, at the same beam energy, while the Fermi momenta in the nucleus smear out the distribution. The black dashed line shows the actual invariant mass distribution of N* resonances from these initial nucleon+nucleon scatterings, which is shifted because the outgoing particles have finite momenta. The red solid line depicts the invariant mass distribution of all N* resonances which decay during the systems whole evolution. Sub-figure (a) presents results of very peripheral collisions of Ca$+$Ca at $E_{\mathrm{lab}}= 1.76$ A GeV, while sub-figure (b) depicts results for head on collisions. In sub-figure (c) we show results for central Be$+$Be collisions at the same beam energy. }\label{f1}
\end{figure*}		

In previous works we have shown that the production of light and strange hadrons
in collisions at low beam energies, at the SIS18, can be consistently described by the UrQMD transport model
where we introduced known cross sections on the strangeness exchange reactions \cite{Graef:2014mra}. We found that 
the production of the $\phi$ meson and $\Xi$ baryon can also consistently described by
the intermediate excitation of a heavy resonance, where the branching ratios were constrained by elementary
reactions like p+p and p+A \cite{Steinheimer:2015sha}. The resulting $\phi$ and $\Xi^-$ yields in collisions of heavier nuclei were consistent with experimental data which indicated that the yields of heavy baryons, containing more than on strange (or anti-strange) quark, are
good measures for the non-equilibrium nature of particle production at this low beam energies.
Even though the successful description of the data was achieved, several questions remained and new ones arose. In this paper we will address these questions which can be outlined as follows:
\begin{enumerate}
\item As the strange particle production occurs as a sub threshold process at the beam energies considered, 
it is quite relevant which processes lead to the accumulation of sufficient energy to produce the heavy hadrons.
In particular we want to distinguish the effects of the Fermi momenta, present in the incoming nuclei, from
secondary interactions of produced hadrons, which can accumulate energy.
\item In the previous paper we have already discussed an additional observable which can be used
as independent confirmation of the processes proposed, i.e. the nuclear transparency of the $\phi$ meson.
In this work we introduce another observable which may be used to confirm our approach. The centrality and system size dependence of the $\Xi^-/\Lambda$ and $\phi/\pi$.
\item Recently the idea that the hadron yields, measured in low energy collisions at the SIS18, can be described by a thermal fit has been put forward \cite{Agakishiev:2015bwu,Agakishiev:2010rs}. Indeed if only the light and single strange stable hadrons are considered one obtains a satisfactory fit. Since the particle yields are also well described with the UrQMD transport model one would expect that the model can also produce a system in apparent chemical equilibrium, with the known exception of the $\Xi$ baryon. It is now interesting to investigate if these apparent equilibrium yields are the result of a long equilibration process in the transport model, following from many multi step processes, or whether hadrons are essentially born into apparent equilibrium after a few steps. We have to note that, because we are dealing with deep sub-threshold processes for strangeness production, single collisions cannot lead to the observed yields of strange hadrons.
\end{enumerate} 

In the following we will answer these questions. The paper is organized such that we first introduce the model used, UrQMD. Then we will investigate the role of Fermi momenta and secondary interactions in the sub threshold production of strange hadrons. In the last part of the paper we will discuss the origin and time dependence of the apparent chemical equilibrium in the model. 

\section{The model}

For the following we will employ the newest version of the UrQMD transport model \cite{Bass:1998ca,Bleicher:1999xi}. The UrQMD model is based on the propagation and scattering of hadrons and hadronic resonances. All the hadron properties, i.e. their masses, quantum numbers, widths as well as scattering cross sections are, where known, taken from the particle data book \cite{Agashe:2014kda}. In recent publications we have presented new features that were implemented and which turned out to be essential for the description of strange particle production below their elementary energy threshold \cite{Graef:2014mra,Steinheimer:2015sha}. These processes are the strangeness exchange reactions $Y + \pi \leftrightarrow N + \overline{K}$ as well as new decay channels for the most massive baryonic resonances included in UrQMD. As we have found in previous publications, the inclusion of such processes allows for a good description of heavy ion collision data at the SIS18 energy regime. In addition, our approach has been shown to describe the absorption cross section of the $\phi$ meson in a cold nuclear environment.
 
It is important to note that in the current version the model does not include any additional medium effects which would change the hadron properties in nuclear collisions and we have not included any hadronic potentials. We believe that is of great importance to first establish a benchmark of particle production before one should discuss possible effects of hadronic long range and multi particle interactions. 
Such effects have been discussed in particular in the context of kaon+nucleon interactions which have been found to influence the resulting kaon and anti-kaon spectra in nuclear collisions \cite{Brown:1991kk,Weise:1996xk,Waas:1996tw,Lutz:2003id,Fuchs:2005zg,Hartnack:2011cn,Schaffner:1996kv,Wisniewski:2001dk,Forster:2007qk,Benabderrahmane:2008qs,Agakishiev:2010zw,Zinyuk:2014zor,Aichelin:1986ss,Shor:1989nz,Hartnack:1993bq,Fang:1994cm,Li:1994cu,Li:1994vy,Mosel:1992rb,Miskowiec:1994vj,Cassing:1996xx,Bratkovskaya:1997pj,Hartnack:2001zs,Hartnack:2005tr,Cabrera:2014lca}. In order to quantify the strength of the hadronic interactions one needs to establish a baseline from elementary reactions, and its systematic uncertainties. 
A dedicated study concerning the kaon spectra at SIS18 energies, in the UrQMD model, is underway and will be discussed in the future. In the present paper however, we will focus on the effects and changes on measured particle yields which occur due to multiple scatterings in nuclear collisions, as compared to elementary collisions. 
Thus we will discuss effects like secondary scatterings as well as Fermi momenta on particle production.

\section{Results}

First we investigate the production mechanisms of the most heavy N* resonances, in order to interpret our previous results on $\phi$ and $\Xi$ production. As we have discussed before there are two mechanisms in nuclear collisions, in contrast to elementary p+p collisions, to produce particles with larger mass than the available invariant energy in the p+p collision. One of these mechanisms stems from the fact that nucleons inside a nucleus obtain Fermi momenta and therefore individual collisions of target and projectile participants may have a larger invariant mass than given by the beam energy. We can define the maximally available invariant energy, for production of a N* in such a collision of nucleons of the projectile and target nucleus, as the total invariant energy minus the second nucleon mass.

\begin{figure}[t]	
\includegraphics[width=0.5\textwidth]{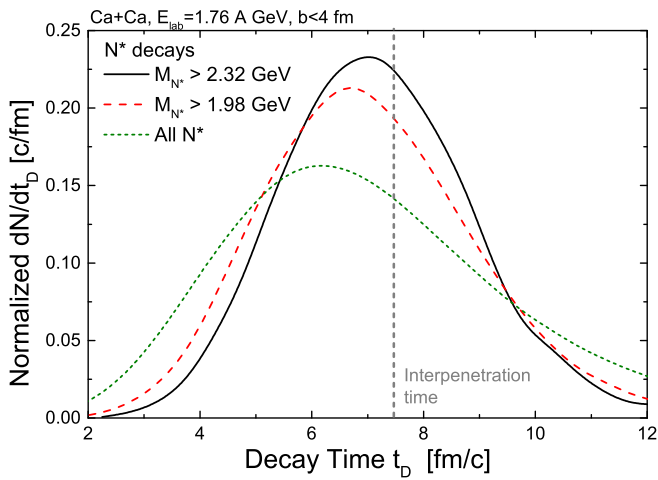}	
\caption{[Color online] Distribution of the decay times of N$*$ resonances (green short dashed line). We also show the distributions of resonances with masses larger than $\mathrm{M}_{\mathrm{N}*}$. Apparently high mass resonances decay at a later time (roughly by 1 fm/c) than lighter resonances, as they have to be created in secondary collisions an therefore appear at a later stage of the collision.} \label{f2}
\end{figure}		

The distribution of this totally available energy, in several different colliding systems, at the same beam energy of $E_{\mathrm{lab}}= 1.76$ A GeV is shown in figures \ref{f1} as blue short dashed lines. The peak of this distribution obviously is located at the beam energy (in the CM frame of the collision) minus the nucleon mass. However we also observe a significant smear of the distribution which can originate only from the Fermi momenta of the nucleons, as we consider only collisions of target and projectile nucleons which have not yet scattered. As the vertical grey lines indicate the energy required to produce a N*, with sufficient mass to subsequently create a $\phi$ or $\Xi$, one can clearly observe, that the Fermi momenta alone are not sufficient to lift the energy above their required production threshold. This becomes even more clear as we also depict the actual invariant mass distribution if N* resonances from such initial scatterings of the incoming nucleons (as black dashed line). The distribution is shifted to lower masses by almost 200 MeV. This shift can be explained from the fact, that the resonances produced in the original scatterings do not obtain the maximally available mass, but acquire a finite momentum which then lowers the mass of the nucleus. Furthermore the production probability is in fact a result of the single N* production cross section which serves as input in the UrQMD model, hence we also observe a certain structure in the mass distribution. 
  
If we now compare this primary production to the mass distribution from all decaying N* resonances (during the whole time evolution of the system), we observe that this distribution is significantly changed. The late time regeneration of resonances introduces a significant peak at the low mass end of the distribution, as well as significantly enhances the high mass tail which is responsible for $\phi$ and $\Xi$ production. It is therefore clear that the heavy resonances which then may decay into heavy multi strange particles are created in secondary interactions where primarily produced resonant states serve as an energy reservoir. It is interesting to note that this accumulation of energy through subsequent inelastic collisions has the same net effect as the introduction of multi particle (mean field) potentials, where the available and necessary energy to produce a heavy hadron is shared and accumulated by the interacting system of hadrons.

The idea that secondary interactions serve to accumulate energy for the production of heavy hadronic states is supported by an observation regarding the decay times of N* resonances, as depicted in figure \ref{f2}. Usually one expects that heavier resonances have smaller lifetime, but as is shown in the figure we observe that heavier N* resonances appear to decay, on average, at a later time than the lighter states. The only reason for this can be that they are actually created at a later time, as they have to go through at least 2 or more collisions to accumulate enough mass. Only at the very late stage of the collision, for $t>10$ fm/c, we observe again an increased decay of lighter resonances, which are essentially from late stage regeneration of light resonances from meson + baryon interactions.

\begin{figure}[t]	
\includegraphics[width=0.5\textwidth]{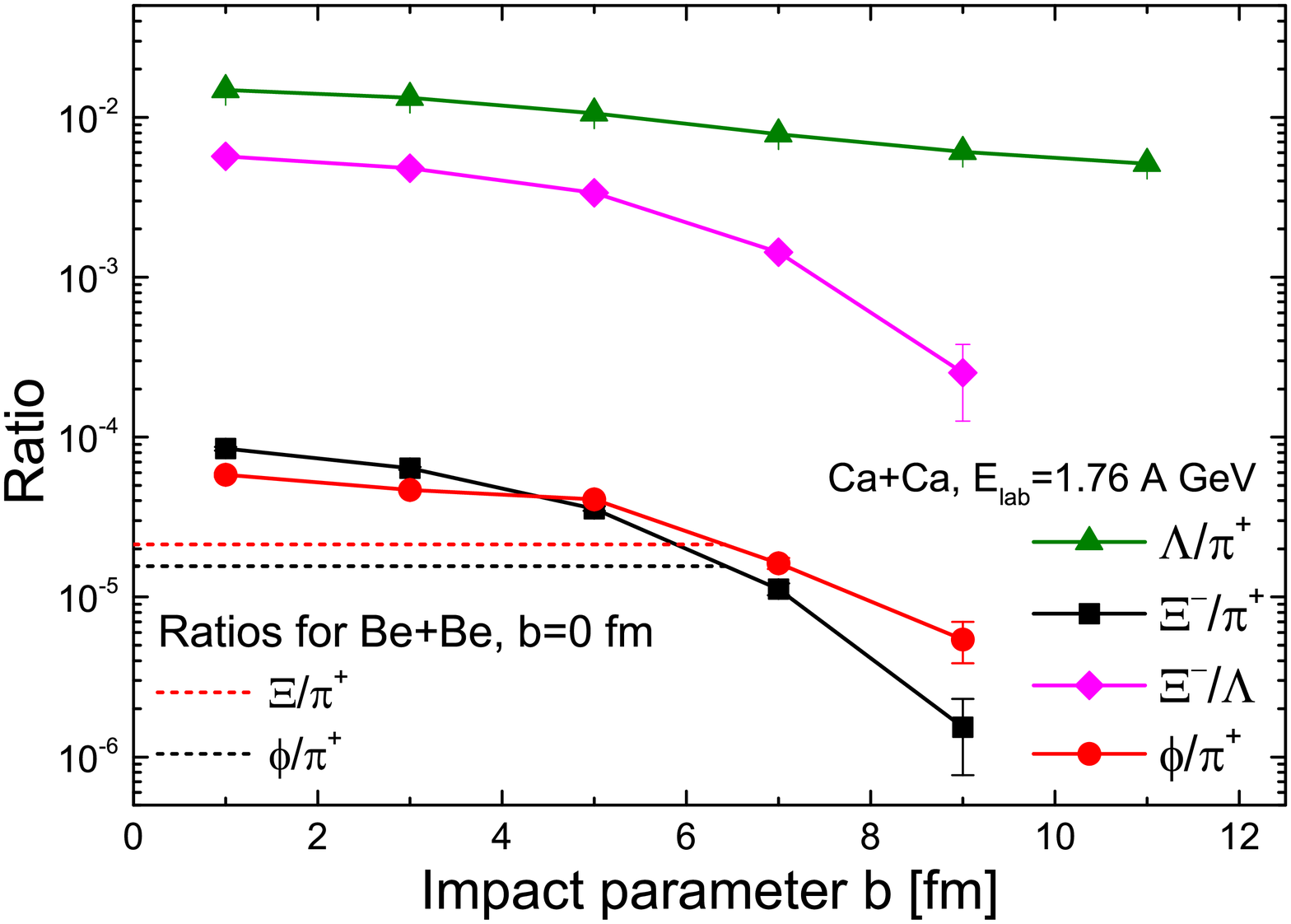}	
\caption{[Color online] Centrality dependence of several strange particle ratios for collisions of
Ca+Ca at a fixed target beam energy of $E_{\mathrm{lab}}= 1.76$A GeV, as simulated with UrQMD. A strong centrality dependence for the $\Xi^-$ and $\phi$ are observed as they can only be created in secondary reactions. We compare the results from Ca+Ca with central collisions (b=0 fm) of the smaller system Be+Be (dashed lines). The ratios in the Be+Be collision system are consistent with peripheral Ca+Ca collisions, indicating a universal $\mathrm{N}_{\mathrm{part}}$ scaling as expected for multiple step processes.}\label{f3}
\end{figure}		

Since we established that the production of the heavy resonance states originates from secondary interactions with 
incoming target and projectile nucleons, we can assume that there should be a strong dependence of the availability of such states on the abundance of participating nucleons in the reactions. In other words we expect a strong dependence of the multi strange particle yields on collision centrality and/or system size. We therefore investigated the dependence of several strange particle ratios as a function of centrality, determined by the impact parameter $b$, in the UrQMD model, for collisions of Ca+Ca at a beam energy of $E_{\mathrm{lab}}= 1.76$A GeV. The results are depicted in figure \ref{f3}. As expected the ratios of multi strange hadrons show a much stronger centrality dependence as the single strange. Even the $\Xi^-/\Lambda$ ratio has a very strong centrality dependence which 
may serve as a signal for the $\Xi^-$ production through secondary reactions of incoming nuclei. 
We have also simulated very central head on collisions of Be+Be at the same beam energy. The resulting ratios are shown as dashed horizontal lines in figure \ref{f3}. For both, the $\Xi^-/\pi^+$ and $\phi/\pi^+$ ratio
from the Be+Be collision correspond to a Ca+Ca collision at a centrality which has the same number of participants as a $b=0$ Be+Be collision. Again this points to a very strong dependence of multi strange hadron production on the number of participants and therefore on the number of possible secondary interaction partners.

\begin{figure}[t]	
\includegraphics[width=0.5\textwidth]{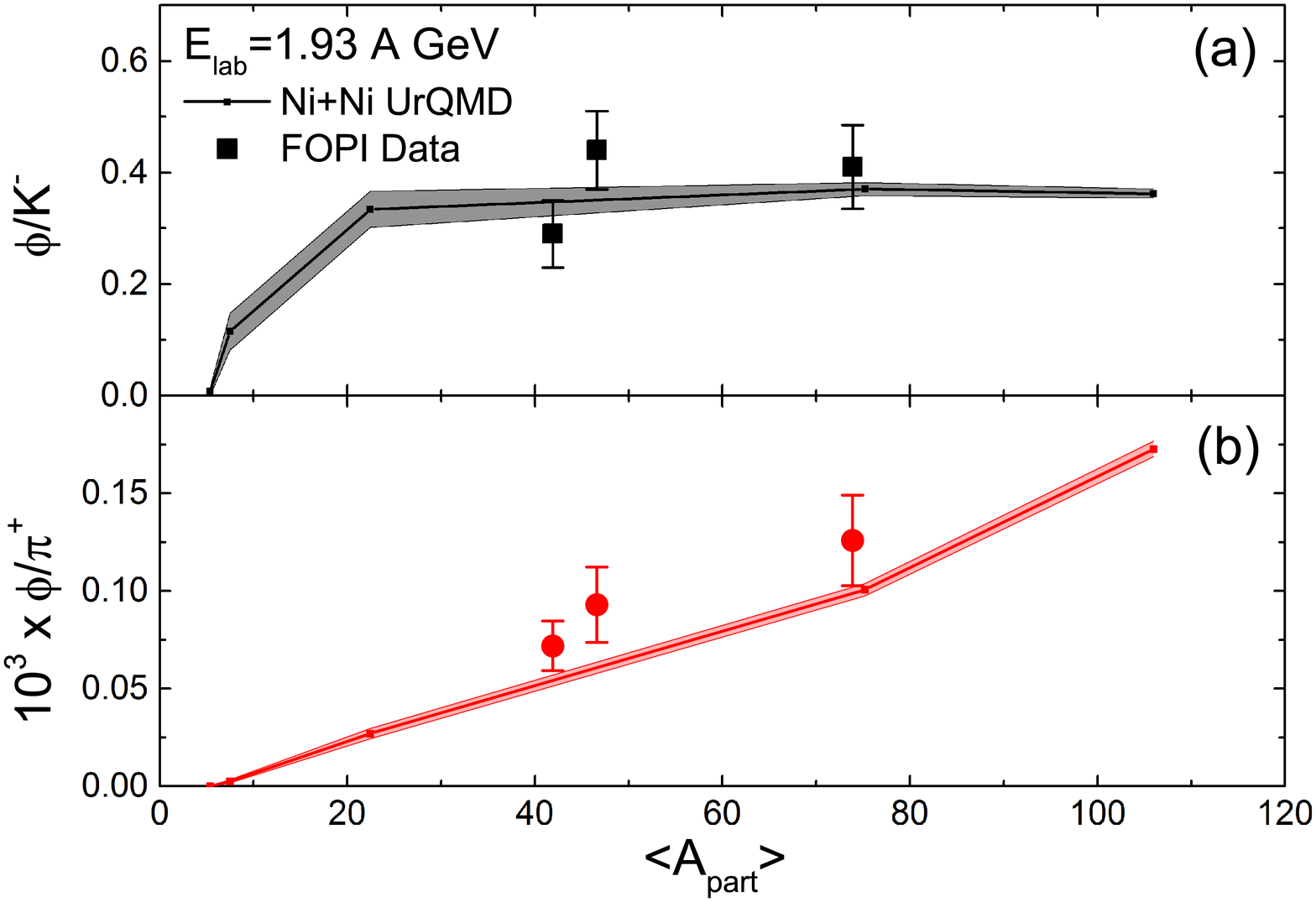}	
\caption{[Color online] Centrality dependence of the $\phi$/$K^-$ and $\phi$/$\pi^+$ ratio in collisions of Ni+Ni nuclei at a beam energy of $E_{\mathrm{lab}}= 1.93$ A GeV. We compare the UrQMD results (lines) with data from the FOPI collaboration \cite{Piasecki:2016tkf}. The model results well describe the measured centrality dependence.}\label{f3b}
\end{figure}

Recently the centrality dependence of sub threshold $\phi$ production has been published by the FOPI collaboration. In figure \ref{f3b} we compare our UrQMD results on $\phi$ production in collisions of Ni+Ni nuclei at a fixed target beam energy of $E_{\mathrm{lab}}=1.93$ A GeV with the data presented in \cite{Piasecki:2016tkf}. The model agrees well with the data and shows a strong centrality dependence of the $\phi$/$\pi^+$ ratio, due to the increasing number of secondary interactions in more central collisions. We understand this result as another validation of our approach.

\subsection{Apparent Equilibration}

An interesting finding from experiments at SIS18 was that the particle abundances of stable hadrons appear to be well described by a fit to an equilibrated chemical ensemble of hadrons. Only few exceptions, for example the production yields of $\Xi$ baryons and some resonance yields, from this finding were observed. Here the question arises why a system, which consists essentially of hadrons, can equilibrate in the short time scales that are involved in such collisions. On the other hand we have found that the hadronic transport model UrQMD does give a reasonable description of all observed particle ratios and therefore can also be described as producing hadrons in apparent equilibrium. The advantage of the dynamical microscopic model is however, that we can in detail investigate the particle production and find out at which point during the interaction the equilibration (if it indeed occurs at all) takes place and how long the systems remains in such a state.

\begin{figure}[t]	
\includegraphics[width=0.5\textwidth]{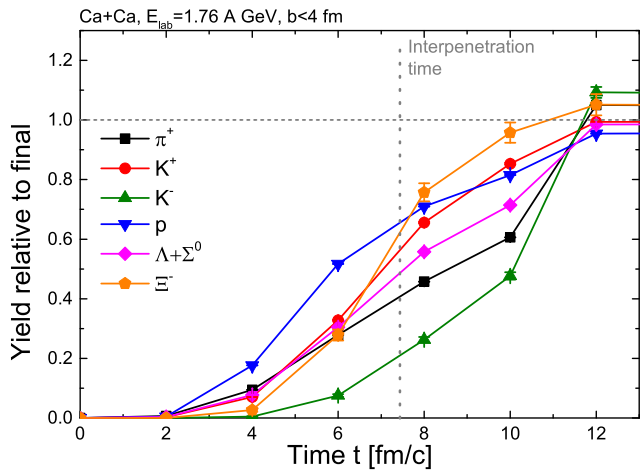}	
\caption{[Color online] Time evolution of the number of several stable particles in central (b$<$4 fm) collisions of Ca+Ca at a fixed target beam energy of $E_{\mathrm{lab}}= 1.76$A GeV. The number extracted refers to particles without resonance decays. It can be seen that the number increases slowly for pions and $K^-$, as they are usually 'hidden' in resonances.}\label{f4}
\end{figure}		

To do so we first extracted the time dependence of stable particle yields from a sample of UrQMD simulations of Ca+Ca collisions at a beam energy of $E_{\mathrm{lab}}= 1.76$ A GeV. The results for the different particle species are shown in figure \ref{f4}, where all yields are normalized to their final (at a time T=100 fm/c) value. We observe that different particles approach their final value at a different rate. For example the $\Xi^-$ approaches it's asymptotic value quickly, while the pions and especially negatively charged Kaons only slowly appear in the collision zone. The reason for this behavior is simple to understand. The negatively charged kaons come, to a relevant degree from the decay of long lived $\phi$ mesons and the pions are usually bound inside resonances for a long time, thus their late appearance.  
 
\begin{figure}[t]	
\includegraphics[width=0.5\textwidth]{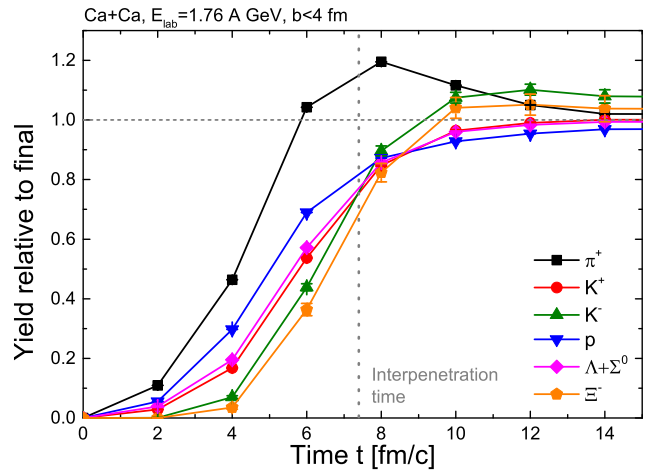}	
\caption{[Color online]  Time evolution of the number of several stable particles in central (b$<$4 fm) collisions of Ca+Ca at a fixed target beam energy of $E_{\mathrm{lab}}= 1.76$A GeV. The number extracted refers to particles after resonance decays. Even if instant resonance decays are taken into account, the number of final stable particles increases differently for different particle species. The pion number in actually overshoots the final value and saturates later, while the $\Xi^-$ number increases slower as it is produced in secondary interactions.}\label{f5}
\end{figure}

Here we have to point out that the time dependence we have observed so far is not useful if we want to study the approach of the particle yields to equilibrium. Usually one considers stable particle yields after resonance decays. Therefore we show again the time dependence of the stable particle yields for the aforementioned collisions, but this time forcing all resonances to decay at the particular time of output. The results are shown in figure \ref{f5}. Now the picture has changed considerably. If we force all resonances to decay at the point of output, the number of pion appears to be rising quickly and even overshoots the final value. This can be understood as a result of the pions usually being 'hidden' in resonances during the most of the evolution. At the late stage these resonances can also be absorbed and therefore the pion yields starts to decrease. It is noteworthy that the maximum of the pion number coincides with the point of maximal compression, i.e. the point in time of the largest central density. This point marks the end of the interpenetration phase during which the energy of the incoming target and projectile nucleons is deposited in the system.
All other particle numbers increase as a function of the collision time. After considering the immediate decays, the $\Xi^-$ yield even grows the slowest which supports the idea that it actually is not produced in the first binary scatterings, but in secondary collisions. 

The yield relative to final for some particles, mainly pions and $K^-$ is apparently
larger than one at late times (note that we did not include any data points for times between
14 and 100 fm/c). For the pions this is due to late time absorption 
reactions of the type $N+N^* \rightarrow N+N$ which effectively lowers the final pion yield.
For the strange $K^-$, the source of the absorption is the strangeness exchange reaction 
$N+K^- \rightarrow Y + \pi$ which also increases the hyperon yields, as observed in figure \ref{f5}.
The strangeness exchange cross section is large at small momenta and therefore shows 
still an effect at late times where relative momenta are small. 
 
\begin{figure}[t]	
\includegraphics[width=0.5\textwidth]{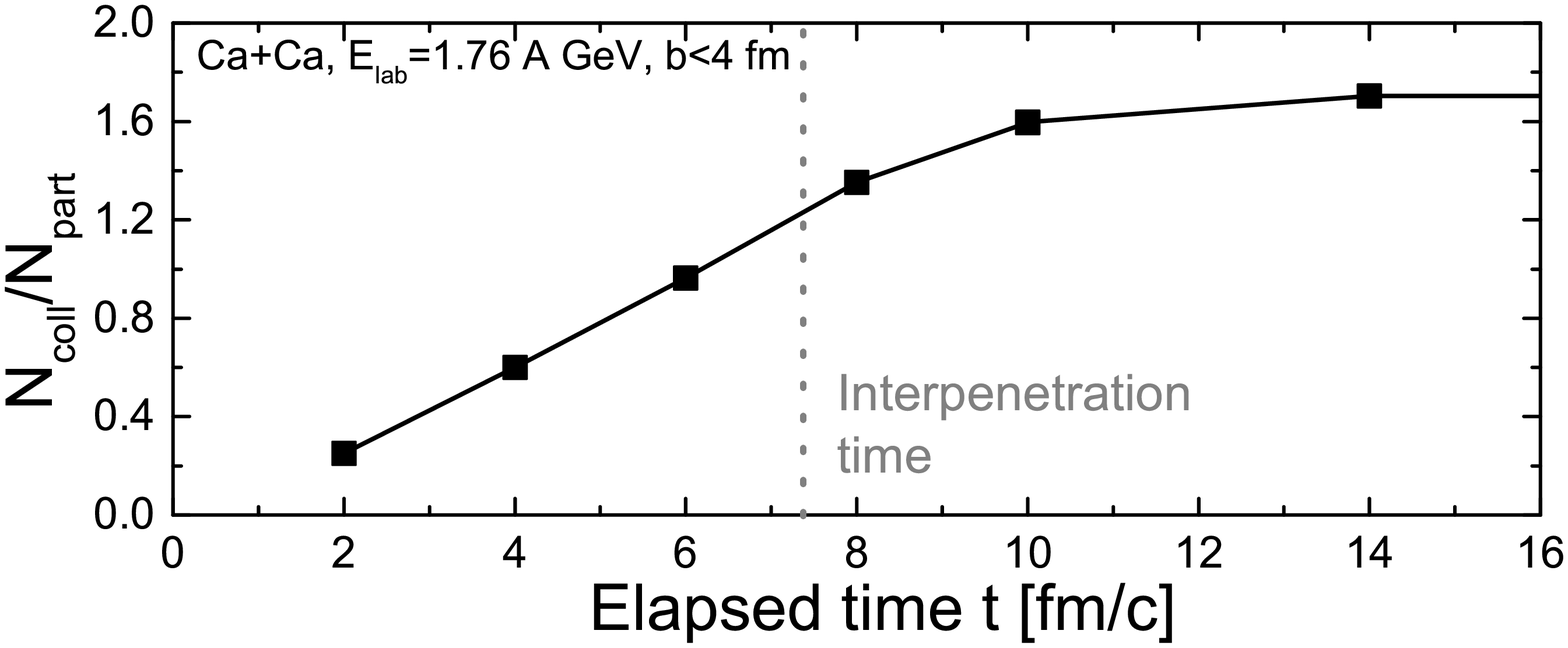}	
\caption{Time evolution of the number of inelastic collisions per participant nucleon in central (b$<$4 fm) collisions of Ca+Ca at a fixed target beam energy of $E_{\mathrm{lab}}= 1.76$A GeV. Apparently the number of inelastic collisions saturates quickly after the interpenetration time, indicating that the chemical composition of the system is fixed rather early and is essentially only changed by secondary collisions during the penetration phase of the two nuclei.}\label{f6}
\end{figure}

It is now interesting to quantify the degree of equilibration that appears to be reached by these secondary reactions.
In particular we would like to know how many inelastic collisions are required to bring the system to a state which appears to be
very close to chemical equilibrium. In figure \ref{f6} we present the time evolution of the average number of 
inelastic collisions, which has occurred until the time $t$, scaled with the number of participants at that time.
As expected the number of inelastic collisions increases with time. The rate at which it increases is nearly constant until shortly after the interpenetration time, after which it slowly saturates. It is expected, that during the 
interpenetration of the nuclei, the most inelastic collisions occur. It is however surprising, that the total 
number of collisions per nucleons remains small during the whole evolution. Even after interpenetration
only 1-2 inelastic collisions per participant nucleon have occurred. This number appears small, however in \cite{Cassing:1990dr}
is was argued that already 3 collisions per participant may be sufficient to achieve thermal equilibrium.

In order to quantify how well the system produced resembles a hadron gas in chemical equilibrium we have
made a fit of the calculated multiplicities presented in figure \ref{f5} to the statistical hadronization 
model (SHM). To cross-check the fits, we have employed two well known and tested codes: the THERMUS package 
\cite{Wheaton:2004qb} and the code used in the analysis~\cite{florence}.

The SHM and its applications to heavy ion collisions has been described in detail elsewhere \cite{Cleymans:1992zc,BraunMunzinger:1995bp,Becattini:1997uf,Becattini:1997ii,BraunMunzinger:2001ip,
Baran:2003nm,Florkowski:2001fp,Cleymans:2004pp,Becattini:2003wp,Vovchenko:2015cbk}. The basic 
assumption is that hadronization occurs in a state of chemical equilibrium at some common value
of the local temperature and chemical potentials. Hence, the primary hadronic multiplicities are completely
determined by few thermodynamical parameters. For large enough volumes, the multiplicities can be 
calculated with the grand-canonical ensemble according to the formula: 
\begin{eqnarray}\label{shmeq}
 n_j &=& (2S_j+1) \frac{V}{(2\pi)^3} \nonumber \\ 
     &\times & \int d^3 p \frac{1}{\exp[E_j/T - \mu_B B_j - \mu_S S_j -\mu_Q Q_j]\pm1}
\end{eqnarray}
where $T$ is the temperature, $\mu_B,\mu_S$,$\mu_Q$ are the baryon, strangeness and chemical potentials,
$V$ is a global volume, $S_j$ is the spin of the hadron $j$ and $E_{j}=\sqrt{m_{j}^{2}+p^2_j}$ its energy;
the upper sign is for fermions, the lower for bosons. Once the primary yields are obtained with (\ref{shmeq}), the
feed-down contribution is calculated by using the experimentally measured branching ratios.

At SIS energies, however, the grand-canonical formula is not a good approximation for the calculation 
of yields. Particularly the strange particles are so few that the effect of exact conservation of
strangeness (the so-called canonical suppression) cannot be neglected. The formula (\ref{shmeq}) has
to be modified to take this effect into account and the statistical ensemble to be used is the so-called
strangeness-canonical ensemble \cite{Becattini:2003wp}. In the limit of Boltzmann statistics, which 
is an excellent approximation for all strange particles, this results in a multiplicative factor
which approaches, in the limit of large volumes and temperatures, the grand canonical fugacity
$\exp[-\mu_S S_j]$.

Furthermore, it is known that particles with strange valence quarks are not at full chemical 
equilibrium, that is they deviate from the prediction of (\ref{shmeq}). The undersaturation of 
strangeness phase space is implemented differently in the aforementioned codes: in
the code of ref.~\cite{florence}, an ad hoc strangeness suppression factor $\gamma_S$ powered to the 
number of valence strange quarks, multiplying the exponentials in (\ref{shmeq}) is introduced; 
instead, in THERMUS, vanishing strangeness is enforced within a reduced volume (strangeness 
correlation volume), enhancing the effect of canonical suppression. It should be pointed out that 
these two methods of extra suppression lead to different predictions of the hidden strangeness 
particle production, like the $\phi$ \cite{Agakishiev:2015bwu,Agakishiev:2010rs}, which are not 
affected by strangeness canonical suppression. 

\begin{figure}[t]	
\includegraphics[width=0.5\textwidth]{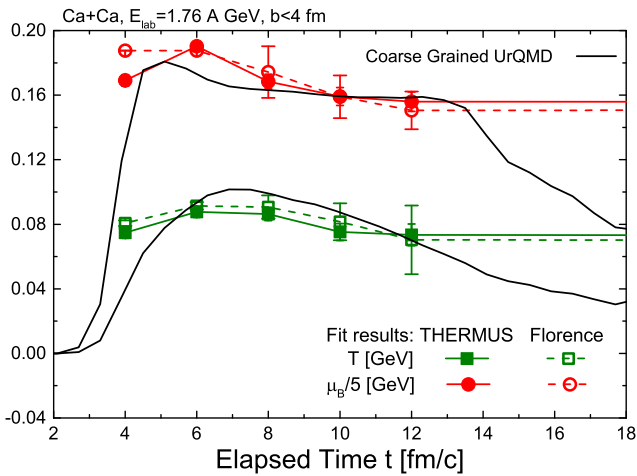}	
\caption{[Color online] Results of two SHM fits to the time dependent stable particle yields obtained with UrQMD simulations of central (b$<$4 fm) collisions of Ca+Ca at a fixed target beam energy of $E_{\mathrm{lab}}= 1.76$A GeV.
The SHM fits were performed with the THERMUS package (solid lines) and the Florence code (dashed lines). We show the temperature (green squares) and baryo chemical potential (red circles) as function of time. The thermal parameters already saturate around 10 fm/c. We compare the fit results with the thermal parameters obtained from a coarse graining approach (black lines, see text).}\label{f7}
\end{figure}

The results of the fits can be seen in figure \ref{f7} as solid (THERMUS) and dashed (Florence) lines.
The $\Xi^-$ hyperon was excluded from the fit and the errors used in the $\chi^2$ were basically the same
as in the experimental measurements. Here we show the extracted temperature and baryon chemical 
potential as function of time. Both SHM approaches give similar results on the extracted thermal 
parameters. As found in the fits to experimental data, the only effect of an inclusion of the $\Xi^-$
in the fit data set is a dramatic increase of the minimum $\chi^2$ whereas the best-fit parameters 
are almost unaffected. The obtained thermodynamical parameters show a clear time dependence. They 
increase during the early stage of the collision, reaching their maximum roughly around the time 
of interpenetration, and then decrease slowly, reaching their final value at a time of about 13 fm/c.

The interpenetration time corresponds approximately to the point of largest temperature (7 fm/c) 
and as shown in \cite{Endres:2015fna} also to the point of largest baryon density. Since the nuclei have 
a finite surface thickness the point of largest compression is already reached shortly before interpenetration,
as the number on incoming nuclei decreases just before that due to the finite thickness.
However, the chemical potential has its maximum
at an even earlier time (5-6 fm/c) which does not coincide with the maximum in the temperature.
In fact there is no reason why the point of highest density should coincide with
the highest chemical potential, as there is still significant entropy generation 
which decreases the ratio $\mu_B/T$, thus decreasing $\mu_B$ at higher baryon density.

\begin{figure}[t]	
\includegraphics[width=0.5\textwidth]{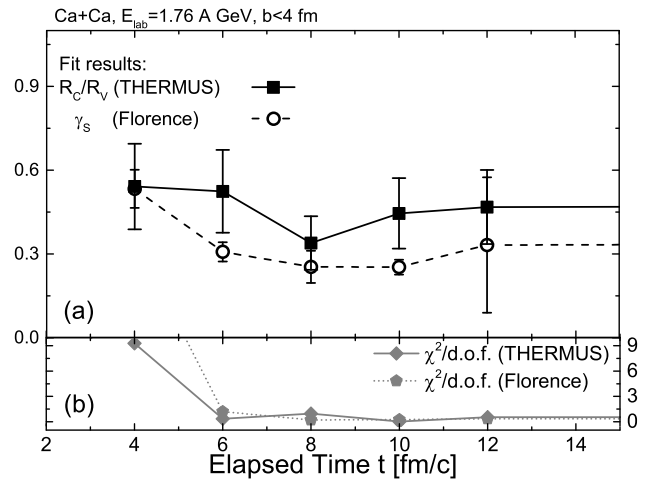}	
\caption{[Color online] Results of a SHM fit to the time dependent stable particle yields obtained with UrQMD simulations of central (b$<$4 fm) collisions of Ca+Ca at a fixed target beam energy of $E_{\mathrm{lab}}= 1.76$A GeV.
The SHM fits were performed with the THERMUS package (solid lines) and the Florence code (dashed lines). We show in figure (a) the ratio of the strangeness canonical radius over the system radius (black squares) and the $\gamma_s$ parameter (black circles). Both quantities have a similar systematic behavior and can be seen as equivalent. Most interestingly, we show in figure (b) the quality of the fit as function of time. Except for the very early times, where no good fit could be obtained, the fit quality is always satisfactory.
}\label{f8}
\end{figure}		

Instead of using a statistical model analysis one can apply another method to extract the temperature and chemical potential, of the produced fireball, during its evolution. Such a method, the coarse graining, has been first applied in \cite{Endres:2015fna} to the system and beam energy under investigation. In the coarse graining approach one uses a large sample of UrQMD events to calculate the average local energy-momentum tensor. One then makes the assumption of local thermal equilibrium to extract energy and baryon number densities in the center of the collision region. These can then be translated to the intensive quantities, temperature and chemical potential, by the use of a hadronic equation of state (for details see \cite{Endres:2015fna}). The values of the temperature and chemical potential that were found in the coarse graining approach are depicted as black solid lines in figure \ref{f7}. There is an astonishing agreement of these values with the ones found by our thermal model analysis. In the time between 5 and 13 fm/c both approaches give essentially the same results. Only at later time, when the chemical abundances freeze out we start to observe expected differences. This coincidence is very surprising and non-trivial, as both approaches are very different and it is interesting that they give similar results. One would expect such a coincidence if indeed the system was in local equilibrium, which we already found to be unlikely due to the few scatterings that occur during the evolution.
It is important to note that the coarse graining results are calculated for a central cell of approximately
1 fm length. Since the comparison of both methods is done for
a rather 'small' colliding system (Ca+Ca) this central cell comprises 
a significant portion of the total system. Furthermore the (average) local Temperature
in early nuclear collisions usually shows only a small gradient until it reaches the systems boundary 
where it drops rapidly (see e.g.\cite{Steinheimer:2007iy}). We also observe that the temperature from the coarse grained
calculations is slightly larger than our average, less than 5 MeV which may be a result
of the fact that we compare our full system with the central cell of the coarse grained approach.
For completeness we refer the reader to a slightly different approach of extracting the thermal parameters from UrQMD with very similar results, presented in \cite{Galatyuk:2015pkq}.
  
The parameters which represent the under saturation of strangeness in the collisions, namely the strangeness canonical radius $R_C$ and strangeness suppression factor $\gamma_S$ are shown in figure \ref{f8} as function of time. Here again both fits are essentially compatible and the canonical radius and strangeness suppression show the systems systematic and quantitative trend. Consequently the only difference in using either parameter is their effect on 
hidden strangeness hadrons, such as the $\phi$ meson which are suppressed by $\gamma_S$ but not by $R_C$. 

In the bottom part of figure \ref{f8} we compare the fit quality, as represented by the $\chi^2$ per degree of freedom, for the two thermal models. Both show an astonishingly good fit quality for times larger than 5 fm/c, in fact better than the ones observed for experimental data. Only in the very first time step, a time when only very few inelastic collisions have occurred, a bad fit quality is observed. Again it appears unlikely that such a good fit quality could be the result of many scatterings leading to an
equilibration process, but is the result of only very few (1-2) resonance excitations and their decays.

\section{Discussion and Conclusions}

We have presented results on strange particle production in the UrQMD transport model at SIS18 energy nuclear collisions. To summarize the most important findings of this work:

\begin{itemize}
\item We have found that the Fermi momenta of the nucleons in the nuclei are not sufficient to lift the threshold constraint on the production
of massive multi strange particles like the $\phi$ and $\Xi^-$, in collisions of Ar+KCl at $E_{\mathrm{lab}} = 1.76$ A GeV. The production yield of these hadrons is therefore very sensitive to multi step production processes and in particular the excitation of heavy baryonic resonances through secondary interactions. We have proposed that such a process leads to a very strong centrality dependence of the $\phi/\pi$ and $\Xi/ \Lambda$ ratios. This is confirmed by the FOPI data.
\item We have calculated the time dependence of stable hadron yields, assuming immediate decays, and used these as an input for a SHM analysis. We found that the hadron yields can be well described by an hadronic gas in chemical equilibrium during most of the systems bulk evolution. At very early times $t\le 5$ fm/c only few initial binary collisions have occurred and a significant fraction of these primary reactions are elastic N+N scattering. Consequently the resulting particle yields cannot be well described within the thermal model ( i.e. yielding a bad fit quality). Once however secondary reactions are taking place, leading to the excitation of numerous baryonic resonances, the corresponding particle yields can be well described with the thermal model, i.e. they resemble yields of a hadronic system in chemical equilibrium. This apparent equilibrium persists until a time of $t\ge 13$ fm/c after which the inelastic collisions cease and the hadronic yields are frozen. It appears that only very few inelastic collisions are required to create a systems which appears to be in chemical equilibrium, which explains why for example the pion spectra appear thermal, even after a very short time, as found in \cite{Galatyuk:2015pkq}. This may also explain why light particle yields in nuclear and even elementary collisions can be well described by a statistical model fit. On the other hand it shows that the production rate of rare and very heavy hadrons is very sensitive to the non-equilibrium tails of the mass distribution of the resonances created in these secondary collisions, therefore serving as signals for the non-equilibrium nature of particle production.
\item We found that the values of the extensive quantities $T$ and $\mu_B$ obtained with the SHM fit agree remarkably well with the values found with a coarse graining approach. Both methods of defining $T$ and $\mu_B$ are not necessarily related. It is important to note that the apparent state of equilibrium is reached already after very few (1-2) inelastic collisions, essentially after the excitation and decay of one or two resonances. The masses of these resonances have been found to be smeared mainly by the Fermi momenta present in the nuclei as well as few secondary interactions. After the isotropic decay of the resonance, which are according to measured branching ratios, the resulting system resembles a hadronic gas near local equilibrium. The microscopic processes which lead to this equilibration are not, as sometimes assumed many consecutive hadronic scatterings, but rather follow from the excitation and decay of heavy resonances that occurs right after the first generation of binary nucleon-nucleon collisions. These resonances decay in accordance with the various phase space weights of the hadronic decay channels (Fermis "golden rule"). This process creates final states of maximum entropy, susceptible to an equilibrium interpretation, which is inherent in the statistical canonical model. 
\end{itemize}
To summarize: the final hadrons (in our approach) do not result from successive generations of binary collisions, spreading the system uniformly in phase space, but from the decay process of resonances, that "gives birth" to final hadrons in equilibrium of species.

\section{Acknowledgments}
We would like to thank J\"org Aichelin for inspiring discussions.
This work was supported by GSI and the Hessian initiative for excellence (LOEWE) through the Helmholtz International Center for FAIR (HIC for FAIR). The computational resources were provided by the LOEWE Frankfurt Center for Scientific Computing (LOEWE-CSC).


\end{document}